\pdfoutput=1
\documentclass[useAMS, usenatbib]{mnras}
\usepackage[]{natbib}
\usepackage[T1]{fontenc}
\usepackage{aecompl}
\usepackage{rotating}
\usepackage{stmaryrd}
\usepackage{aas_macros}
\usepackage{times}
\usepackage{graphicx}
\usepackage{amsmath}
\usepackage{amssymb}
\usepackage{amsfonts}
\usepackage{placeins}

\def\gsim { \lower .75ex \hbox{$\sim$} \llap{\raise .27ex \hbox{$>$}}}
\def\lsim { \lower .75ex \hbox{$\sim$} \llap{\raise .27ex \hbox{$<$}}}
\newcommand{\eagle}{{\sc eagle}}

\begin{document}
\newcounter{MYtempeqncnt}
\graphicspath{{figs/}} 

\title[Measuring the properties of Lyman-$\alpha$ absorbers]{Measuring the temperature and profiles of Lyman-$\alpha$ absorbers}

\author[A. Garzilli, T. Theuns and J. Schaye]  {Antonella Garzilli$^{1}$\thanks{E-mail:
    garzilli@nbi.ku.dk}, 
  Tom Theuns$^{2}$ and Joop Schaye$^{3}$\\
  $^1$ Discovery Center, Niels Bohr Institute, Copenhagen University, Blegdamsvej 17, DK-2100 Copenhagen, Denmark\\
  $^2$ Institute for Computational Cosmology, Department of Physics, Durham University, DH1 3LE Durham, UK \\
  $^3$ Leiden Observatory, Leiden University, PO Box 9513, 2300 RA
  Leiden, The Netherlands
} \date{Accepted --. Received --; in original form --}

\maketitle

\begin{abstract}
The distribution of the absorption line broadening observed in the
Ly$\alpha$ forest carries information about the temperature, $T$, and
widths, $\lambda_{\rm F}$, of the filaments in the intergalactic
medium (IGM), and the background hydrogen photo-ionization rate,
$\Gamma_{\rm HI}$. In this work, we present and test a new method for
inferring $T$ and $\lambda_{\rm F}$ and $\Gamma_{\rm HI}$ from
combining the distribution of the absorption line broadening and the
median flux. The method accounts for any underlying degeneracies.  We
apply our method to mock spectra from the reference model of the EAGLE
cosmological simulation, and we demonstrate that we are able to
reconstruct the IGM properties.
\end{abstract}

\begin{keywords}
intergalactic medium -- quasars: absorption lines --  large scale
structure of Universe -- methods: data analysis
\end{keywords}

\section{Introduction}\label{sec:intro}
In the $\Lambda$CDM model of cosmology the Universe emerges from
inflation in a quasi-homogeneous state, with small fluctuations in the
density field of matter. From these initial conditions, the Universe
evolves to its current state and becomes populated with structures
such as galaxies and galaxy clusters.  Most of the baryons do not
reside in these dense structures, but in a diffuse medium that fills
intergalactic space, called the intergalactic medium (IGM), that is
organized in a network of sheets and filaments. The chemical
composition of the IGM is mostly primordial, with a minor component of
metals, produced by stars and likely injected in the IGM by galactic
winds and outflows, for a review, see
e.g.~\citet{rauch1998,meiksin2009}. Although \cite{cantalupo2014} have
reported that the IGM can be observed in emission, it has mainly been
observed in absorption, in the spectra from distant and bright
sources, such as quasars. The lack of the Gunn-Perterson trough
\citep{gunn1965} since $z\sim 5$ implies that the IGM is in a highly
ionized state.  According to the current understanding of structures
formation, the IGM is photo-ionized and photo-heated by an
hydrogen-ionizing radiation background (UVB) originating from galaxies
and quasars \citep[e.g.][]{haardt1996,Madau:2015cga}. Recently the UVB
has been measured at $z=0$ from H$\alpha$ fluorescence
\citep{fumagalli2017} or at low redshift, $z<0.5$, from the study of
the IGM \citep{Gaikwad:2016trx, Gaikwad:2017gwb, Viel:2016cub,
  Khaire:2018fqp}.

Hence, the IGM is observable through the absorption of light emitted
by distant bright objects, through the Ly$\alpha$ forest, which is the
collection of Ly$\alpha$ absorption lines. The Ly$\alpha$ forest is
also fluctuating Gunn-Peterson absorption, because the absorption
traces the fluctuations in the underlining neutral hydrogen density
field. The widths of the lines in the Ly$\alpha$ forest are determined
by the clustering of the absorbers and their temperature. For a review
we refer the interested reader to \citet{meiksin2009}.  To gain
information on the timing of reionization and the nature of the
responsible sources, it is important to determine the IGM
temperature. Moreover, the IGM has been used as an indirect probe of
dark matter, and to investigate the free-streaming length of dark
matter \citep{Seljak:2006qw}. Recently, there has been some attempt to
constrain the nature of dark matter with high redshift quasar spectra
\citep{viel2013}. Nevertheless, these latest studies suffer from
uncertainties in the IGM temperature, the IGM temperature is an
astrophysical bias in the study of the nature of dark matter at the
smallest scales \citep{Garzilli:2015iwa}, and from the smallness of
the quasar sample analyzed \citep{Garzilli:2018jqh}. Another motivation
for measuring the temperature of the IGM is the study of the second
reionization of helium, that is known to be completed at $z\simeq
2.7$. The second ionization of helium happens at energy $E=54.4\,{\rm
  eV}$, that is four times larger than the energy required to ionize
hydrogen and about twice larger then the energy for ionizing the first
level of helium, hence much harder sources than the ones responsible
for hydrogen reionization are required for second helium reionization.

Many groups have tried to measure the IGM temperature with different
methods, using Voigt-profile fitting
\citep{schaye1999,Schaye:1999vr,Ricotti:1999hx,McDonald:2000nn,bolton2014,
  Rudie:2012mx,hiss2017}, studying the flux PDF
\citep{theuns2000,Bolton:2007xi, Viel:2009ak, Calura:2012qq,
  Garzilli:2012gy}, the flux power spectrum, wavelet analysis and
curvature method \citep{theuns2000,Theuns:2000va,Zaldarriaga:2000mz,
  Viel:2005ha,Lidz:2009ca,Becker:2010cu,Garzilli:2012gy}. The details
of the results from the different methods vary, but there is a general
consensus that in the redshift interval between 2 and 4, $5000\,{\rm
  K}<T_0<30000\,{\rm K}$, where $\Delta=\rho/\left<\rho\right>$ is the
overdensity and $T_0=T(\Delta=1)$ is the temperature of the IGM at the
cosmic mean density.

The width of the structures causing the absorption has been measured
for the first time from pairs of quasars by \citet{rorai2017}. The
intensity of the photo-ionizing background, $\Gamma_{\rm HI}$, has
been measured by previous authors
\citep{rauch1997,mcdonald2001,meiksin2004,bolton2005,kirkman2005,faucher2008},
but always assuming a specific thermal history for the IGM, (see
\citet{fumagalli2017} for a measurement of the ultraviolet background
at low redshift that is independent of the IGM). Over the redshift interval
between 2 and 4, the measurements  agree in finding $2\times
10^{-13}\,{\rm s^{-1}}\leq\Gamma_{\rm HI}\leq 2\times 10^{-12}\,{\rm
  s^{-1}}$.

As already pointed out by \citet{huignedinzhang1997}, there are at
least two distinct physical effects that contribute to the minimum
line broadening in the Ly$\alpha$ forest\footnote{There is an
  additional contribution from the finite resolution of the
  spectrograph.}: the first is the thermal Doppler effect, that is set
by the temperature of the IGM, the second is the extent of the
filaments in the IGM -- the filaments are not virialized structures
and there is a contribution of the differential Hubble flow across the
absorbers
\citep{gnedinhui1998,theuns2000,schaye2001,Desjacques:2004xy,Peeples:2009uj,rorai2013,Garzilli:2015bha,Kulkarni:2015fga}.
The simulations of \cite{schaye1999} and \cite{Ricotti:1999hx} showed
that the minimum line broadening as a function of overdensity can be
approximated by a power-law. In \cite{Garzilli:2015bha}, we
demonstrated that, under the hypothesis of a photo-ionized IGM, the
lower envelope of the line broadening distribution is a convex
function of the baryon density, and hence of the neutral hydrogen
column density. We introduced an analytical description for the
minimum amount of line broadening present in the Ly$\alpha$ forest. In
this same work, we introduced the `peak decomposition' of the neutral
hydrogen optical depth, which differs from the standard Voigt profile
fitting of the spectra described by eg. \citet{carswell1987}.

In this work, we present a new method for measuring the properties of
the IGM from quasar absorption spectra, considering only the
Ly$\alpha$ forest for each quasar spectrum. We develop the method
using mock sightlines extracted from hydrodynamical simulations. We
carry out the measurements using the distribution of Doppler
parameters measured as described in Paper 1 \citep{Garzilli:2015bha}.
We also combine the distribution of absorption line broadening with
the median of the flux, and we obtain the constraints on the
IGM properties that are the main result of this work.

This paper is organized as follows. In Section~\ref{sec:simulations},
we describe the reference EAGLE simulation, from which we have
extracted the mock spectra. In Section~\ref{sec:formulas}, we discuss
the analytical description of the line broadening we use in this
method, and the modifications with respect to the equations presented
in \cite{Garzilli:2015bha}. In Section~\ref{sec:method}, we discuss
the reconstruction of the line broadening in the case of spectra with
noise.  In Section~\ref{sec:results} we discuss the ability of our
method to correctly constrain the IGM parameters from quasar spectra
with noise. In Section~\ref{sec:conclusions}, we present our
conclusions.  In Appendix~\ref{app:vpfit}, we compare with Voigt
profile fitting, which has been used widely in previous works.  In
Appendix~\ref{app:eta50}, we have explicitly shown that our method is
robust respect to the calibration with numerical simulations.  In
Appendix~\ref{app:lowsn}, we will show that our conclusions do not
change in the case of lower signal to noise spectra: in this case we
merely obtain larger error bars on the estimated parameters.

\section{Simulations}\label{sec:simulations}
\subsection{The \eagle\ simulations and the $T/\Delta$ relation}\label{sec:TDR}
In this paper, we use the 25~cMpc (co-moving Mpc) high-resolution
reference simulation of the \eagle\ suite \citep{Schaye15,
  Crain15,mcalpine2016}, labelled \lq L0025N0752\rq\ in table~2 of
\cite{Schaye15}. The simulation is based on the \cite{planck2014}
values of the cosmological parameters, and the initial baryonic
particle mass is $1.81\times 10^6$~M$_\odot$. This cosmological
smoothed particle hydrodynamics (SPH) simulation is performed using
the {\sc gadget-3} incarnation of the code described by
\cite{Springel05}, with modifications to the hydrodynamics algorithm
referred to as {\sc anarchy} (described in the Appendix~A of
\cite{Schaye15}, see also \citealt{Schaller15}). The reference model
incorporates a set of sub-grid models to account for unresolved
physics, which include star formation, energy feedback and mass loss
feedback from stars, black halo formation, accretion and merging, and
thermal feedback from accreting black holes. The parameters that
encode these sub-grid models are calibrated to observations of $z\sim
0$ galaxies, namely the galaxy stellar mass function, galaxy sizes,
and the stellar mass - black holes mass relation, as described in
detail by \cite{Crain15}.

The simulation also accounts for photo-heating and radiative cooling
in the presence of the imposed background of UV, X-ray and CMB
radiation described by \cite{Haardt01}, using the interpolation tables
computed by \cite{wiersma2009a}. The optically-thin limit is assumed in
these simulations.

Photo-heating and radiative cooling, adiabatic cooling due to the
expansion of the Universe, and shocks from structure formation and
feedback, result in a range of temperatures for cosmic gas at any
given density.  However, the majority of the gas follows a
single-valued relation - or better, follows a well-defined relation
between the temperature and the density, the so-called
temperature-density relation (TDR for short). We will indicate the
over-density of the gas with $\Delta=\rho/\langle\rho\rangle$, where
$\rho$ is the density of the gas and $\langle \rho\rangle$ the cosmic
mean density, whereas we will indicate the temperature of the gas at
any given over-density, $\Delta$, with $T(\Delta)$.  At $\Delta\le 3$,
the TDR is set by the interplay between photo-heating and adiabatic
cooling, resulting in an approximately power law relation
$T=T_0\,\Delta^{\gamma-1}$ \citep{Hui97, Theuns98,
  Sanderbeck:2015bba}. When the temperature of the cosmic gas is
increased rapidly by photo-heating, as happens during hydrogen
reionisation, the slope is $\gamma\approx 1$, whereas asymptotically
long after reionisation, it becomes $\gamma\approx 1+1/1.7\approx
1.6$, as discussed by \cite{Hui97} and \cite{Theuns98}. The fact that
$\gamma$ in this limiting case is close to that of the adiabatic index
of a mono-atomic gas, $\gamma=5/3$, is a coincidence. During the
second reionization of helium, that we know to be completed by
$z=2.7$, the picture is a bit different, because different sources of
ionizing radiation are involved. In fact, the second level of helium
requires a ionization energy $E=54.4\,{\rm keV}$, that is four times
larger than the ionization energy of hydrogen. While early galaxies
are thought to be the source of hydrogen ionization, the sources of
second helium reionization are thought to be quasars. Because of the
different distribution of the sources and hardness of their spectra,
the temperature configuration in density is also different, giving a
power-law with $\gamma\sim 1.3$ and much larger scatter
\citep{mcquinn2009,Puchwein:2014zsa,Puchwein:2018arm,
  Gaikwad:2018ojv}.

At higher overdensity, $T$ is set by the balance between photo-heating
and radiative cooling. This causes a gentle turn-over in the
$T-\Delta$ relation around $\Delta=3$ at redshift $z>1$ (above
$\Delta\approx 30$ at $z=0.5$).  In this work we will consider the
simple case that the $T-\Delta$ relation is a power-law, and we leave
the investigation of more physically motivated $T-\Delta$ relations
for future work.

\subsection{Mock sightlines}
We compute mock sightlines from the \eagle\ simulation. We begin by
sampling the simulation volume with sightlines parallel to its
$z$-axis, using pixels of velocity width $W_v=1.4$~km~s$^{-1}$, which
is small enough to resolve any absorption features. We next use the
interpolation tables from \cite{wiersma2009b} to compute the neutral
hydrogen fraction for each SPH particle in the optically-thin limit,
taking the cosmic gas to be photo-ionised at the rate calculated by
\cite{Haardt01}. We then compute the contribution of each gas particle
to the spectrum (or better, to the temperature, density and line of
sight velocities along the spectrum) by integrating a kernel over each
pixel, calculating the \ion{H}{I} density, and the \ion{H}{I}-weighted
temperature and peculiar velocity. This is similar to the algorithm
described in the Appendix of \cite{Theuns98}, except that here we
integrate over each pixel rather than evaluating the kernel at the
centre of the pixel. Kernel integration is much simplified by using a
Gaussian kernel rather than the M4-spline used in {\sc gadget}, and we
do so as described by \cite{Altay13}.

Each pixel generates a Gaussian absorption profile of the form
\begin{eqnarray}
\tau &=& \tau_0\,\exp(-v^2/b_{\rm T}^2)\label{eq:gaussiantau}\\ b_{\rm
T}^2 &=&
\left({2k_{\rm B}\,T\over m_{\rm H}}\right)\\ \tau_0 &=& \sigma_0
     {c\over \pi^{1/2}\,b_{\rm T}}\,N_{\rm HI}\\ \sigma_0 &=&
     \left({3\pi\,\sigma_T\over 8}\right)^{1/2}\lambda_0\,f\,,
\label{eq:spec}
\end{eqnarray}
where $v=v_{\rm z, pix}-v_{\rm z, part}$ is the velocity difference
between the pixel and the particle in the $z$-direction, and $N_{\rm
  HI}$ is the neutral hydrogen column density of the pixel\footnote{In
  practice we integrate the Gaussian in Eq.~\ref{eq:gaussiantau} over
  a pixel, rather than evaluating it at the pixel centre.}. The
physical constant appearing in these equations are the speed of light,
$c$, Boltzmann's constant, $k_{\rm B}$, the hydrogen mass, $m_{\rm
  H}$, and the Thompson cross section, $\sigma_T$.  For the
Lyman~$\alpha$ transition, the wavelength and $f$-values are taken to
be $\lambda_0=1215.6701$~\AA~ and $f=0.416$, see \cite{menzel1935}.
Since we are analysing cosmic gas at densities around the mean
density, we do not need to use the more accurate Voigt profile. For
more details, we refer the reader to the Appendix A4 in
\citep{theuns1998}.
	
While the simulation is running, we output those particles that
contribute to one hundred randomly chosen sightlines, for every
10~per cent increase in the cosmic expansion factor. This allows us to
account accurately for any redshift evolution in the generated mock
sightlines. 
The computation of the mock sightlines only takes into account the
Ly$\alpha$ transition. We leave the consideration of other transitions
of the Lyman series for future work.  In addition to computing $\tau$,
we record the optical-depth weighted temperature, peculiar velocity,
and overdensity as a function of wavelength. \cite{Garzilli15}
demonstrate (their figure~1) that the relation between optical depth
weighted temperature and density follows that of the actual TDR.

In the next section, we analyse mock sightlines generated with and
without noise. We intend to mimic the properties of some observed
spectra, and for example we consider the properties of spectra
measured with HIRES \citep{1994SPIE.2198..362V}. Hence, mock
sightlines with noise are obtained by convolving the transmission
$\exp(-\tau)$ with a Gaussian profile with full width at half maximum,
rebinning the sightlines into pixels of $W_v=4\,{\rm km\,s^{-1}}$,
$f_{\rm FWHM}$, equal to $6.6$~km~s$^{-1}$ and adding random Gaussian
noise corresponding to a chosen signal-to-noise per pixel at the
continuum, S/N=100 or 30. This ensures that resolution and
signal-to-noise in the mock spectra are comparable to those of
high-quality HIRES or UVES spectra \mbox{\citep{kim2007}}.

\section{Analytical expression for the minimum absorption line broadening}\label{sec:formulas}
In \cite{Garzilli:2015bha} we provided an analytical expression for
the minimum absorption line broadening, $b$, as a function of the
over-density, $\Delta$, associated to the line. Unfortunately $\Delta$
cannot be measured directly from the observed spectra. Hence, here we
derived a relation between $b$ and the central neutral hydrogen
optical depth in an absorption line, $\tau_0$.

We start from the expression of the optical depth as in
\cite{miralda1993}
\begin{eqnarray}
  \tau(u_0) & = &\Sigma \int^{u_{\rm B}}_{u_{\rm A}} \frac{n_{\rm
      HI}}{1+z}\left|\frac{du}{dx}\right|^{-1} \sigma_{\alpha} du
  \nonumber \\
  \sigma_{\alpha} & = & \sigma_0 \frac{c}{b_{\rm T}
    \sqrt{\pi}} e^{-\frac{(u-u_0)^2}{b_{\rm T}^2}} \nonumber \\
  \frac{du}{dx} & = & \frac{H(z)}{1 + z} + \frac{\partial v_{\rm
      pec}}{\partial x}\, ,
\end{eqnarray}
where $n_{\rm HI}$ is the neutral hydrogen number density, $x$ is the
comoving spatial coordinate, $u$ is the velocity along the line of
sight, $u_{\rm A}$ and $u_{\rm B}$ are the extremes of the absorber
expressed in velocity along the line of sight, $\sigma_{\alpha}$ is
the Ly$\alpha$ cross-section profile, $b_{\rm T}=\left( 2 k_{\rm B}
T/m_{\rm H} \right)^{1/2}$ is the thermal Doppler broadening, $T$ is
the temperature of the gas, $v_{\rm pec}$ is peculiar velocity, $z$ is
the mean redshift associated to the absorber and $H(z)$ is the Hubble
parameter at redshift $z$, the summation is made over multiple streams
of $x$ with the same $u$.

As already demonstrated by \cite{theuns2000}, the effect of peculiar
velocities on the line broadening is
\begin{enumerate} 
\item shifting the position of the absorption
lines;
\item narrowing or broadening the absorption lines -- \emph{peculiar
  velocities do not always broaden the absorption lines, as if they
  were a turbulent contribution}.
\end{enumerate}
We have explicitly tested the effect of peculiar velocities on the
absorption line broadening distribution.  We have demonstrated that
they do not affect the its overall shape \citep{Garzilli:2015bha}. For
clarify, we want to make explicit that the turbulent motion is due to
peculiar velocities, hence saying that peculiar velocities are
negligible for our purposes is equivalent to say that turbulence is
negligible. For these reasons, we neglect peculiar velocities and we
assume a Gaussian profile for the neutral hydrogen number density,
$n_{\rm HI}^0 \exp\left\{-(u-u_0)^2/b^2_{\lambda}\right\}$, where
$n^0_{\rm HI}$ is the neutral hydrogen number density at the center of
the line and $b_{\lambda}$ is the width of Gaussian profile of $n_{\rm
  HI}$. Hence, the neutral hydrogen optical depth becomes
\begin{eqnarray}
  \tau(u) & = & \frac{\sigma_0 c n^0_{\rm
        HI}}{H(z)} \frac{b_{\lambda}}{b}e^{-{\frac{(u-u_0)^2}{b^2}} } \, ,
\end{eqnarray}
where we will assume the line broadening to be equal to the `minimal`
line broadening
\begin{eqnarray}
  b\equiv b_{\rm min} = \left(b_{\rm T}^2+b_{\lambda}^2\right)^{1/2}\,
  .
  \label{eq:bminimal}
\end{eqnarray}
The width of the Gaussian profile for $n_{\rm HI}$ can be expressed as
\begin{eqnarray}
  b_{\lambda} & = & \lambda_{\rm F} \frac{H(z)}{2\pi} \, ,
\end{eqnarray}
as in \cite{Garzilli:2015bha}, where $\lambda_{\rm F}$ is the
proper extent of the absorbing structure.  Again as
in \cite{Garzilli:2015bha}, we make the Ansatz that
\begin{eqnarray}
  \lambda_{\rm F} & = & f_{\rm J} \lambda_{\rm J}(\Delta) \, ,
\end{eqnarray}
where $f_{\rm J}$ is a constant that parametrizes the time-dependent
Jeans-smoothing of the gas density profiles \citep{gnedinhui1998}, and
$\lambda_{\rm J}$ is the local Jeans length of an absorber
\citep{schaye2001}. Here we use
\begin{eqnarray}
  \lambda_{\rm J}(\Delta) & = &\pi \Big(\frac{40}{9} \Big)^{1/2}
  \Big(\frac{k_{\rm B}}{m_{\rm H} } \Big)^{1/2} (1+z)^{-3/2} 
  {H_0}^{-1} {\mu}^{-1/2} {\Omega_m}^{-1/2}\nonumber \\
  &\times & T^{1/2} {\Delta}^{-1/2}  \, ,
\end{eqnarray}
where $\mu$ is the mean molecular mass, $\Omega_m$ is the matter
density parameter and $H_0$ is the Hubble constant. In the following
we will indicate with $\lambda_{\rm F}$ the proper extent of
absorbing structure at the  cosmic mean density.

We consider the TDR we have described in
section~\ref{sec:TDR}. Because we consider explicitly the
temperature--density relation, our method takes explicitly into
account the dependency of the temperature with density. We
make explicit how $b$ and the optical depth at the center of the line,
$\tau_0$, depend on $\Delta$, the temperature
at cosmic mean density, $T_0$, the slope of the TDR, $\gamma$, the
proper width of the absorbers of the Ly$\alpha$ forest, $\lambda_{\rm
  F}$, and the intensity of the hydrogen ionizing background,
$\Gamma_{\rm HI}$
\begin{eqnarray}
  \tau_0 & = & \frac{\sigma_0 c n^0_{\rm HI}}{H(z)} 
  \frac{b_{\lambda}}{\sqrt{b_{\rm T}^2 + b_{\lambda}^2}}\label{eq:opticaldepth}\\
  n_{\rm HI}^0 & = &  \alpha_0 \frac{9}{128 \pi^2}
  (m_{\rm H} G)^{-2}  (2-Y)(1-Y) (1+z)^6 {H_0}^4 {\Omega_b}^2 \nonumber \\
  & \times &   {\Gamma_{\rm HI}}^{-1} \left(\frac{T_0}{10^4\,{\rm
      K}}\right)^{-0.76}  \Delta^{2.76-0.76\gamma}\nonumber\\
  \alpha_0&=&4\times 10^{-13} {\rm cm^3 s^{-1}}\, ,  \label{eq:btau0}
\end{eqnarray}
where $\Omega_b$ is the baryon density parameter, $Y$ is the helium
fraction by mass, $G$ is the gravitational constant, $\alpha_0$ is the
recombination constant at $T=10^4\,{\rm K}$. We do not provide an
explicit the relation between $\tau_0$ and $b$, but it can be computed
by inverting numerically Eq.~(\ref{eq:btau0}) with respect to $\Delta$.

We intend to quantify the minimal line broadening as a function of the
density and compare with our analytical description of the
broadening. As we have already discussed the density is not directly
observable in the observed spectra. Hence, we resort to quantify the
minimal line broadening as a function of the optical depth, so that we
can compare with Eq.~(\ref{eq:btau0}). In the following, we will show
that in the presence of noise we cannot reconstruct the minimal line
broadening over a wide range of optical depth, but we can reconstruct
the median of the line broadening as a function of the optical depth.

\begin{figure}
  \includegraphics[width=\columnwidth]{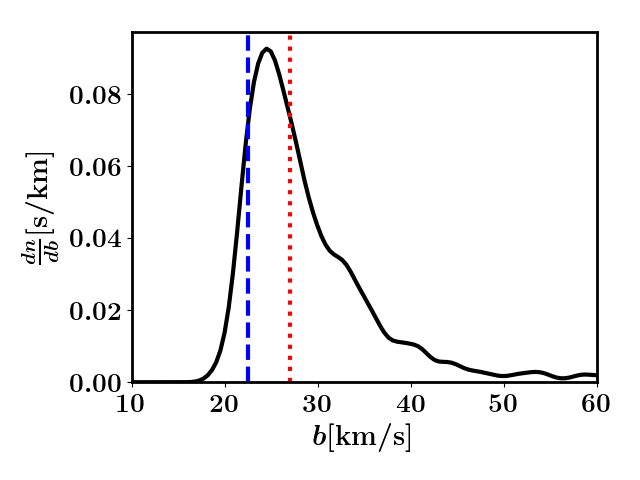} \caption{The line
    broadening probability distribution for lines with
    $0.8\leq\Delta\leq 1.2$, estimated by Kernel Density. We show the
    10th and the 50th percentiles of the distribution as the dashed
    blue and dotted red lines, respectively.} \label{fig:perc}
\end{figure}

We now make a comparison with our fiducial simulation. We measure
Doppler parameters by applying the peak identification method from
paper I \citep{Garzilli:2015bha}, not to be confused with the
traditional Voigt profile fitting -- a comparison between the two
methods is given in Appendix~\ref{app:vpfit}. Our peak identification
method has been formulated to be applicable to spectra without
noise. While the traditional Voigt profile fitting method consider the
flux in the spectrum, in our peak identification method we consider
the optical depth of the spectrum as a function of velocity. Then, we
identify the minima of the optical depth, each stretch of spectrum
between two consecutive minima is considered a `peak', and the maximum
optical depth within the peak is the `central optical depth', and it
is considered to be an estimator for $\tau_0$. Because we consider
spectra without noise, we can compute the second derivative of the
optical depth with respect to velocity at the maximum of each peak in
the spectra, $\tau_0^{''}=\phantom{\lvert}\frac{d^2
  \tau}{dv^2}{\rvert}_0$. For each identified peak in the spectrum, we
can associate a line broadening, $b$, from the central optical depth
and the second derivative, $b^2=-2\tau_0/\tau_0^{''}$.  In
Figure~\ref{fig:perc}, we show the probability density function of the
line broadening for the interval $0.9\leq \Delta\leq 1.1$ (absorbers
around the cosmic mean density), and we show the 10th and 50th
percentiles of the line broadening distribution. The number of lines
decreases rapidly when $b\leq {\rm mode}(b)$, this implies that the
10th percentiles of the probability density function can be used to
approximate the absolute lower limit of $b$.

\begin{figure*}
  \includegraphics[width=\textwidth]{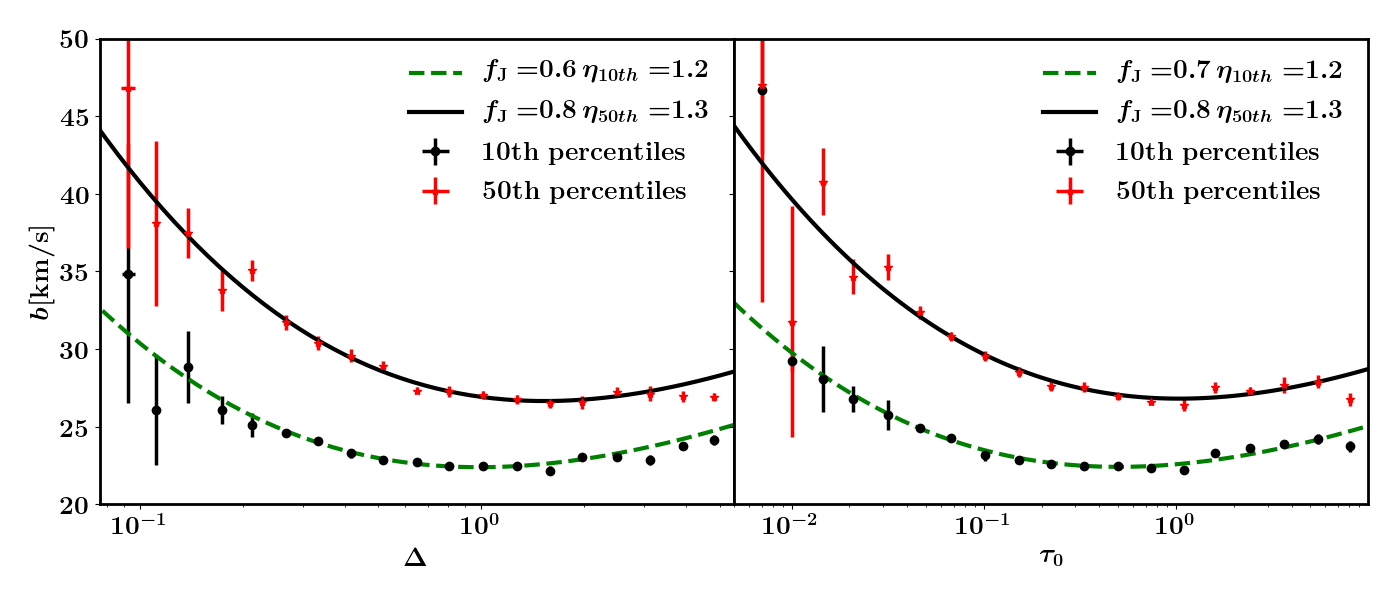} \caption{Percentiles of the
    line broadening, $b$, from 100 noiseless mock spectra versus the
    peak density contrast, $\Delta$ (left panel), and the peak neutral
    hydrogen optical depth, $\tau_0$ (right panel), in the redshift
    interval $2.9\leq z\leq 3.0$. The black dots (red stars) are the
    10th (50th) percentiles of the $b$-distribution in the mock
    spectra. The error bars represent the 1-$\sigma$ uncertainty on
    the percentiles.  In the left panel, the green dashed (solid
    black) line is a fit of our expression for the line broadening,
    Eq.(~\ref{eq:bperc}), to the 10th (50th) percentiles of the
    $b$-distribution divided in logarithmically and equally spaced
    intervals of $\Delta$. The values of $T_0$ and $\gamma$ are the
    result of the fit to the temperature-density relation, the values
    of $f_{\rm J}$ and $\eta$ are the result of the fit of the
    $b$-$\Delta$ relation. In the right panel, we have presented the
    analogous fit to the percentiles of the $b$-distribution as a
    function of $\tau_0$. In the transition from $b$-$\Delta$ space to
    $b$-$\tau_0$, the fitted values of $f_{\rm J}$ agree to within
    15\%. Our analytical expression for the line broadening correctly
    describes both the 10th and the 50th percentiles of line width
    distribution in our reference simulation, except for high values
    of $\tau_0$, where the downturn in $b$ is due to the onset of
    radiative cooling.} \label{fig:btau0}
\end{figure*}

In Figure~\ref{fig:btau0}, we compare the distribution of the line
broadening in the plane $b$-$\Delta$ and in the plane $b$-$\tau_0$, to
highlight their similarity, the absorbing lines are being binned in
$\Delta$ (or $\tau_0$). The error bars on the 10th and 50th
percentiles of the $b$-distribution are computed by bootstrapping the
lines of sight, rather than the absorption lines themselves.
The minimum line broadening is a difficult quantity to measure. 
It is possible in noiseless spectra to measure an arbitrary
percentile of the distribution of the line broadening, but that can
only approximate the minimum line broadening. For this reason, we
adapt Eq.~(\ref{eq:bminimal}), which we have written for the minimum line
broadening, to the case of a generic percentile of the line broadening
distribution, $b_{\rm perc}$
\begin{eqnarray}
 b_{\rm perc}^2 & = & \eta_{\rm perc}^2 (b_{T}^2+b_{\lambda}^2)\, ,
\label{eq:bperc}
\end{eqnarray}
where $\eta_{\rm perc}$ is a constant that will depend on the chosen
percentile of the $b$-distribution. This constant must be determined
from simulations, an incorrect calibration will implies a systematic
effect on the reconstruction of the IGM parameters. The impact of
varying $\eta_{\rm perc}$ is discussed in
\ref{app:eta50}. Eq.~(\ref{eq:bperc}) describes well both the 10th and
the 50th percentiles of the absorption line broadening distribution up
to $\tau_0\sim 10$. Nevertheless, the values of $f_{\rm J}$ can differ
in the fits to the $b$-$\Delta$ distribution or to $b$-$\tau_0$
distribution. The discrepancy in $f_{\rm J}$ between these two
distribution can be up to $\sim15\%$. For $\tau_0\gtrapprox 10$
radiative cooling becomes relevant, although the precise value of
$\tau_0$ for which this occurs depends on the specific values of
$T_0$, $\gamma$, $\lambda_{\rm F}$ and $\Gamma_{\rm HI}$.  In this
work we only consider the case of power-law TDR, but we intend to
address the problem of a general TDR in a future publication.

\section{Method for reconstructing the absorption line
  broadening}\label{sec:method} 
In this section we will discuss how to
reconstruct the minimum line broadening from spectra with noise, and
then how to estimate $T$ and $\lambda_{\rm F}$. In the presence of
noise and instrumental broadening, we cannot apply directly the
formalism that we have developed in \mbox{\cite{Garzilli:2015bha}} for
measuring the line broadening occurring in the Ly~$\alpha$ forest to
observed quasar spectra. In there the line broadening was estimated
directly from the second derivative of the optical depth from mock
spectra without noise, compare Eq.(18) of \mbox{\citep{Garzilli:2015bha}}. In
spectra with noise, the computation of the second derivative in the
measurement of $b$ is not stable under noise. In order to smooth out
the noise, we first fit the sightlines with a superposition of Voigt
profiles, and then apply the procedure we have already developed for
noiseless sightlines on the spectra reconstructed from their Voigt
profile decompositions, and determine $b$ and $\tau_0$ for each
absorber. We then apply Eq.~(\mbox{\ref{eq:bperc}}).  In this section, we
consider the case of mock sight lines to which we added noise with S/N=100.
We use a sample size of a total 500 sightlines for each considered
redshift interval, each spectrum has a length of $25\,c{\rm Mpc}$. If
we analyze together all the signal coming from bins in redshift of
$\Delta z=0.1$, then for redshifts ranging from $z=4$ to $2$, 500
sightlines correspond to a number of Lyman~$\alpha$ quasar spectra
varying from 180 to 85. This sample size is comparable with current
sample size of observed high resolution and high signal to noise
quasars, for example there are $\sim 500$ quasar spectra collected in
\mbox{\cite{2018MNRAS.tmp.2706M}}.

\subsection{Reconstructing the line broadening in the
  Ly$\alpha$ forest}\label{subsec:percentiles} We attempt to remove
the noise in the mock sightlines, by fitting the Ly$\alpha$ stretch in
the mock sightlines with noise with VPFIT
\mbox{\citep{carswell1987,webb1987}}. The full spectrum flux is divided into
intervals of variable length, between 10 and 15~\AA. We start from the
minimum wavelength in the Ly$\alpha$ stretch, $\lambda_1$, then we
search for the maximum of the flux in the interval
$[\lambda_1+10\,$\AA$,\lambda_1+15\,$\AA$]$, the wavelength
corresponding to the maximum flux will be $\lambda_2$. Then, the
maximum of the flux in the interval
$[\lambda_2+10\,$\AA$,\lambda_2+15\,$\AA$]$ is identified and the
corresponding wavelength will be $\lambda_3$. These maxima do not
always coincide with the continuum, there is always some residual
absorption. This process is repeated until the end of the spectrum
has been reached. In this way the spectrum is subdivided into
intervals of variable length. Each interval of transmitted flux is fit
independently with VPFIT.  The stopping criteria that we have
considered is given by the change of the chi-square, $\Delta\chi^2$,
between iteration steps. If $\chi^2<15$ then the iteration terminates
if $\Delta\chi^2/\chi^2<5\times10^{-4}$, otherwise if $\chi^2>15$ the
iteration terminates if $\Delta\chi^2/\chi^2<5\times10^{-3}$. The flux
is also reconstructed separately for each independent stretch. We do
not attempt to perform a full Voigt-profile fitting of the entire
Ly$\alpha$ forest, because of the long computing time required for
fitting automatically the entire Ly$\alpha$ forest in one batch. We
are only interested in a noiseless reconstruction of the flux in the
minimum $\chi^2$ sense.  On the reconstructed optical depth we apply
the `peak identification' method and estimate the line broadening as
described in \mbox{\cite{Garzilli:2015bha}} for the case of noiseless
sightlines.

\begin{figure*}
  \includegraphics[width=\textwidth]{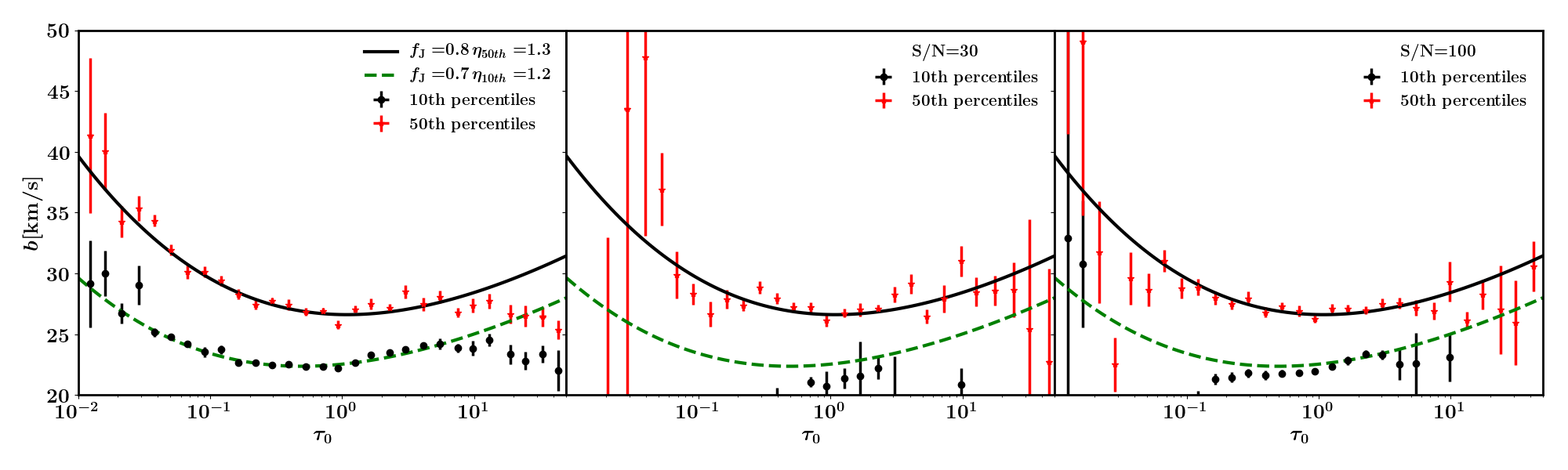}
  \caption{Comparison between the percentiles of the distribution of
    line broadening in the noiseless sightlines, and in the sightlines
    reconstructed with VPFIT, in the case of high and low
    signal-to-noise, for the redshift range $2.9 \leq z \leq3.0$. We
    have considered 500 lines of sight, each of length $25\, {\rm
      cMpc}$ . The left panels correspond to the noiseless case, the
    middle panel corresponds to the sightlines with noise with S/N=30,
    and the right panel to S/N=100. The black dots (red stars) are the
    10th (50th) percentiles of the line broadening as a function of
    the central optical depth, $\tau_0$. The green dashed (black
    solid) line is the result of the fit of Eq.(~\ref{eq:bperc}) to
    the 10th (50th) percentiles of the $b$-distribution, where $T_0$
    and $\gamma$ are inferred from the temperature-density relation,
    and $f_{\rm J}$ and $\eta$ are free parameters in the fit. In the
    case of reconstructed sightlines with VPFIT, the 10th percentiles
    of the $b$-distribution are very poorly reconstructed. In
    contrast, the 50th percentiles of the $b$-distribution are
    reconstructed well over a larger range of $\tau_0$. The robustness
    under reconstruction with VPFIT makes the 50th percentiles more
    suitable for the study of the properties of the IGM.}
  \label{fig:percentiles}
\end{figure*}

In Figure~\ref{fig:percentiles} we show a comparison between the 10th
and 50th percentiles of the $b$-distribution for noiseless sightlines
and for the reconstructed flux for the case of high and low S/N. We
have considered 500 sightlines in the redshift interval $2.9\leq z
\leq 3.0$. We would like to measure the minimum line broadening in the
sightlines, hence ideally we would like to consider the
10th-percentiles of the line broadening. Nevertheless, we can see that
qualitatively the 10th percentiles are not reconstructed very well in
the sightlines with noise. Instead, the 50th percentiles (or medians)
of the line broadening are reconstructed over a larger $\tau_0$
range. The reason is that noise increases the dispersion of the line
broadening distribution at fixed $\Delta$. Hence, while the median of
the distribution is not changed by this added dispersion, the 10th
percentiles are changed.  For this reason, in the following we will
characterize the line broadening by considering the 50th percentiles
of the $b$-distribution, rather than the 10th.

\begin{figure*}
  \includegraphics[width=0.45\textwidth]{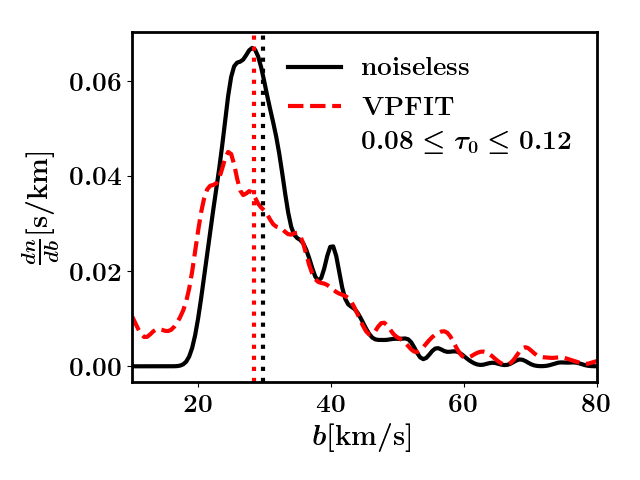}
  \includegraphics[width=0.45\textwidth]{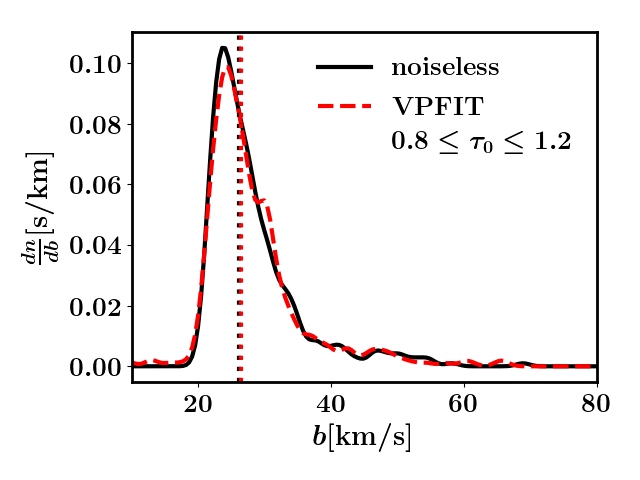}
  \caption{The probability density distribution of the logarithm of
    the line broadening, $\ln b$, for two values of the central
    optical depth, $\tau_0$, for 100 mock sightlines without noise and
    with noise (S/N=100). The probability density function has been
    estimated through kernel density estimate, with Gaussian kernel
    and the kernel bandwith is optimized by cross-validated
    grid-search over a grid. The left panel refers
    to the line broadening corresponding to $0.08\leq \tau_0\leq
    0.12$, the right panel to $0.8\leq\tau_0\leq 1.2$.  The black
    solid line is the PDF in the noiseless sightlines, the red dashed
    line is the one from the sightlines with noise. The black vertical
    line corresponds to the 50th percentile of the $b$-distribution in
    the noiseless sightlines, the red dashed vertical line corresponds
    to the 50th percentile of the $b$-distribution in the sightlines
    with noise. For the interval centered on $\tau_0\sim 0.1$, the
    number density of lines in the noiseless sightlines is $n_{\rm
      noiseless} = 7.1 \times 10^{−4} {\rm s/km}$, whereas the
    number of lines per length in the sightlines with noise is $n_{\rm
      noise}=7.0\times10^{−4}{\rm s/km}$. For the interval
    centered on $\tau_0\sim 1$, the number density of lines in the
    noiseless sightlines is $n_{\rm noiseless} = 9.4 \times 10^{−4}
    {\rm s/km}$, whereas the number of lines per length in the
    sightlines with noise is $n_{\rm noise}=8.4\times10^{−4}{\rm s/km}$.  }
  \label{fig:bpdf}
\end{figure*}

In Figures~\ref{fig:bpdf} we compare the PDF of the $b$-distribution
as found in the noiseless sightlines and in the sightlines with noise,
for two distinct intervals in $\tau_0$. For $0.08\leq \tau_0\leq
0.12$, the PDF of the reconstructed $b$ is much flatter then the PDF
of the noiseless $b$, and the two PDFs do not match each other
well. The number density of lines in the noiseless sightlines is
$n_{\rm noiseless}=9.4\times10^{-4}\,{\rm s\,km}^{-1}$, whereas the
number of lines per length in the sightlines with noise is $n_{\rm
  noise}=8.4\times10^{-4}\,{\rm s\,km^{-1}}$. For $0.8\leq \tau_0\leq
1.2$ the PDFs of the reconstructed and noiseless $b$ are quite
similar, they both exhibit a sharp cut-off for low values of $b$ and a
declining tail for large values of $b$. We conclude that the line
broadening is reconstructed less accurately for smaller values of
$\tau_0$. For $0.8\leq \tau_0\leq 1.2$, the number of lines per length
in the noiseless sightlines is $n_{\rm
  noiseless}=7.1\times10^{-4}\,{\rm s\,km}^{-1}$, whereas the number
of lines per length in the sightlines with noise is $n_{\rm
  noise}=7.0\times10^{-4}\,{\rm s\,km^{-1}}$. We note that the number
density of reconstructed lines refers to both genuine absorption lines
and to fictitious lines originating from noise. We can see that the
number of reconstructed line is comparable to the number of genuine
lines for both interval of the overdensity.  This comparison allows us
to say that we can determine the 50th percentiles of $b$ well for
$\tau_0\sim 1$.

\subsection{Estimation of the IGM parameters}\label{sec:results}
We have considered the estimation of the IGM parameters over a
redshift interval $z\in [2.9, 3.6]$, with a redshift step $\Delta
z=0.1$.

We intend to estimate $T_0$, $\gamma$, $\lambda_{\rm F}$ and
$\Gamma_{\rm HI}$.  $T_0$ and $\gamma$ are the parameters of the TDR,
and they have been the subject of many studies in the past, whereas
the role of $\lambda_{\rm F}$ in setting the line broadening has been
recognized only relatively recently, $\Gamma_{\rm HI}$ is a parameter
that is usually kept fixed to a value. We have decided to vary
$\Gamma_{\rm HI}$ because it affects the optical depth. In fact, when
$\Gamma_{\rm HI}$ is inferred from observations, $T_0$ is assumed. As
an example, we consider the cases of \citep{Becker:2010cu},
\citep{Becker:2013ffa} and \citep{faucher2008}. In those work,
$\Gamma_{\rm HI}$ is fixed to a value, the $T_0$ is measured. From the
measurement of $T_0$, a new measurement of $\Gamma_{\rm HI}$ is
inferred.  

In order to get an estimate of all the parameters that are relevant
for the IGM, we fit the measured 50th percentile of the
$b$-distribution from the sightlines with noise with the model
$b(\tau, T_0, \gamma, \lambda_{\rm F}, \Gamma_{\rm HI})$, our
analytical formula for line broadening, Eq.~(\ref{eq:bperc}). We have
chosen the interval for the reconstructed central optical depth
$\tau_0=[0.1,4]$, in order to exclude the region affected by cooling
and by noise. We divide the central optical depth into equi-spaced
logarithmic intervals, and we compute the median of the line
broadening for each central optical depth bin. We indicate the
resulting collection of central optical depth and median line
broadening values with ${\hat{\tau_i},\hat{b}_i}$, where $i$ is index
that varies on the bins of the central optical depth. We also estimate
the 1-$\sigma$ error on $\hat{b_i}$, $\sigma(\hat{b}_i)$. The errors
are estimated by bootstrapping the lines of sight, rather than the
absorption lines. The theoretical model is indicated with the notation
$b_i(T_0, \gamma, \lambda_{\rm F}, \Gamma_{\rm HI}) = b(\tau_i, T_0,
\gamma, \lambda_{\rm F}, \Gamma_{\rm HI})$. The constant
$\eta_{50th}$, appearing in Eq.~(\ref{eq:bperc}), has been calibrated
using our reference simulation, separately for each redshift interval,
as in Table~\ref{tab:cal}. In Appendix~\ref{app:eta50}, we have
explicitly tested the effect of changing $\eta_{50th}$ by $\pm 5\%$,
and we have demonstrated that our results are unchanged.

\begin{table}
\centering
 \caption{Values of $\eta_{50th}$, appearing in Eq.~(\ref{eq:bperc}),
   calibrated from the noiseless mock spectra from our reference simulation as a function of redshift,
   $z$.}
 \begin{tabular}{l|cc} 
   $z$ & $\eta_{50th}$\\
   \hline
   2.95 & 1.32\\
   3.05 & 1.27\\
   3.15 & 1.25\\
   3.25 & 1.28\\
   3.34 & 1.26\\
   3.45 & 1.26\\
   3.56 & 1.25
 \end{tabular}
\label{tab:cal}
\end{table}

In order to perform the fit, we compute the chi-squared function,
$\chi^2$, defined as 
\begin{eqnarray}
\chi^2 &=& \sum_i \left(b_i(T_0, \gamma,
\lambda_{\rm F}, \Gamma_{\rm HI}) -
\hat{b_i}\right)^2/\sigma(\hat{b_i})^2 \, ,
 \label{eq:chi2}
\end{eqnarray} 
where $b(T_0,\gamma, \lambda_{\rm J}, \Gamma_{\rm HI})$ is the line
broadening as computed from Eq.~(\ref{eq:bperc}), $\hat{b}$ is the
50th percentile of $b$-distribution.  We compute the corresponding
likelihood function by ${\cal L}=\exp(-\chi^2/2)$ and maximize the
likelihood by using Montepython \citep{Audren:2012wb}, CosmoMC
\citep{Lewis:2002ah} and Polychord
\citep{Handley:2015fda,Handley_2015}. We have chosen logarithmic
priors on $T_0$, $\Gamma_{\rm HI}$ and $\lambda_{\rm F}$ and a flat
prior on $\gamma$, which are summarized in Table~\ref{tab:priors}.
\begin{table}
\centering
 \caption{Prior ranges considered for the parameter of the maximum
   likelihood analysis, used for fitting the mock median line
   broadening data to the model given by Eq.~(\ref{eq:bperc}). We have
   chosen logarithmic priors on $T_0$, $\Gamma_{\rm HI}$ and
   $\lambda_{\rm F}$ and a linear prior on $\gamma$. Here and in
   the rest of the paper, $\log$ indicates the logarithm in base 10.}
 \begin{tabular}{l|cc} 
   & min & max\\
   \hline
   $\log(T_0[{\rm K}])$ & 0 & 5 \\ 
   $\gamma$ & 1 & 2 \\
   $\log(\lambda_{\rm F}[{\rm cMpc}]$) & -3 & 3\\
   $\log(\Gamma_{\rm HI}[{\rm s^{-1}}])$  & -13 & -11
 \end{tabular}
\label{tab:priors}
\end{table}

\begin{figure*}
  \includegraphics[width=\columnwidth]{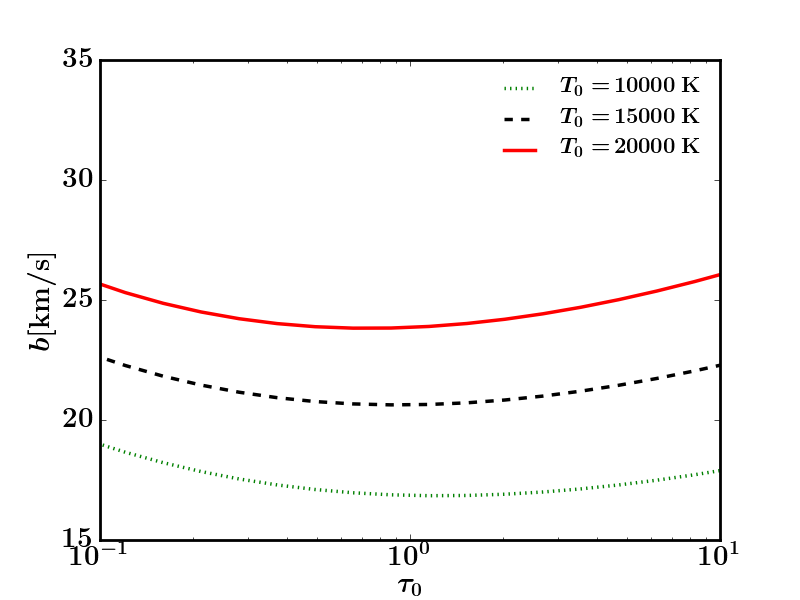}
  \includegraphics[width=\columnwidth]{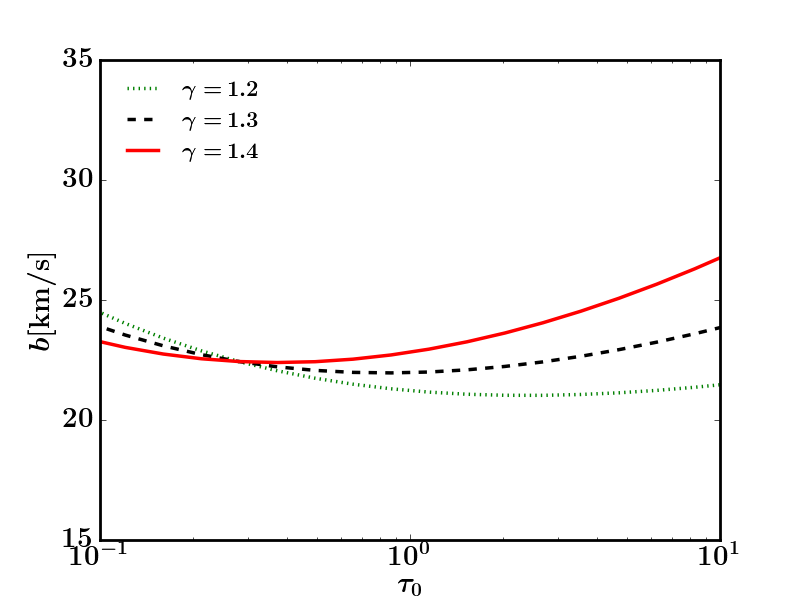}
  \includegraphics[width=\columnwidth]{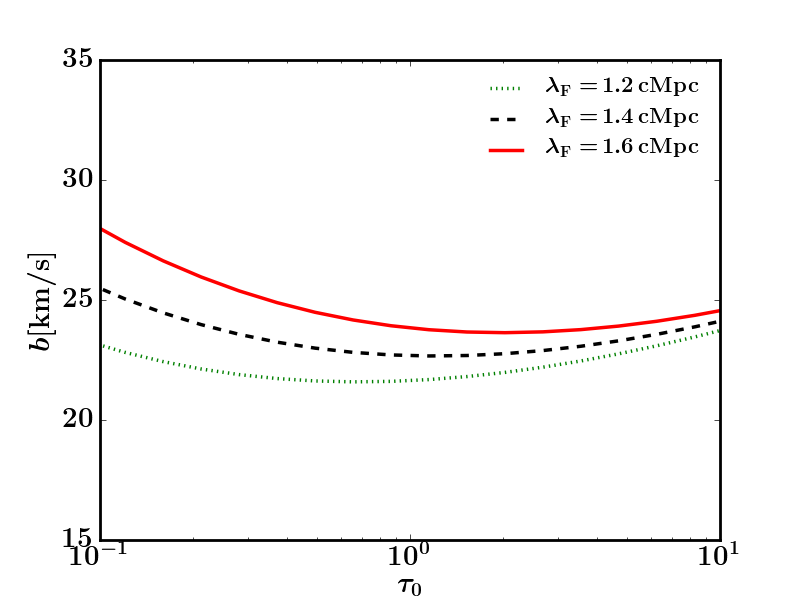}
  \includegraphics[width=\columnwidth]{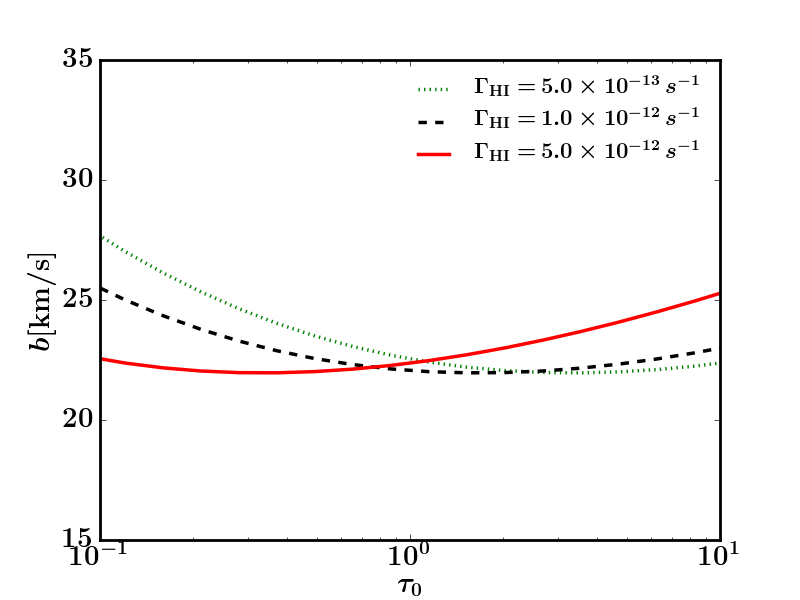}
   \caption{The effect on the line broadening of varying $T_0$,
     $\gamma$, $\lambda_{\rm F}$ and $\Gamma_{\rm HI}$
     independently. The line broadening is computed from
     Eq.~(\ref{eq:bperc}) for $z=3.0$ (with $\eta=1$). The parameters
     that are not shown in the label are fixed to $T_0=17000\,{\rm
       K}$, $\lambda_{\rm F}=1.4\,{\rm cMpc}$, $\gamma=1.3$,
     $\Gamma_{\rm HI}=10^{-12}\,{\rm s^{-1}}$. }
   \label{fig:trends}
\end{figure*}

Before presenting the joint fit of all parameters, we show in
Figure~\ref{fig:trends} how the minimal line broadening is affected by
each parameter independently, using our analytical model of the line
broadening in Eq.~(\ref{eq:bperc}). Changing $T_0$ is almost
equivalent to changing the line broadening by a multiplicative
factor. Changing $\lambda_{\rm F}$ mostly affects the line
broadening at small $\tau_0$. Changing $\gamma$ affects the slope of
the line broadening at all $\tau_0$. Changing $\Gamma_{\rm HI}$ has
the effect of changing the neutral fraction of hydrogen, hence it
affects the $\Delta$-$\tau_0$ relation, and it shifts the position in
the minimum of the $b$-$\tau_0$ relation. The effect of $\Gamma_{\rm
  HI}$ is only to shift the curves of line broadening left to right,
but not up and down.  We can expect some degeneracies between the
estimated parameters in the final analysis: $\gamma$ and $\Gamma_{\rm
  HI}$ appear to be correlated, $T_0$ and $\gamma$ appear to be
anti-correlated, $T_0$ and $\lambda_{\rm F}$ anti-correlated,
$\lambda_{\rm F}$ and $\Gamma_{\rm HI}$ correlated.

\begin{figure*}
  \includegraphics[width=\textwidth]{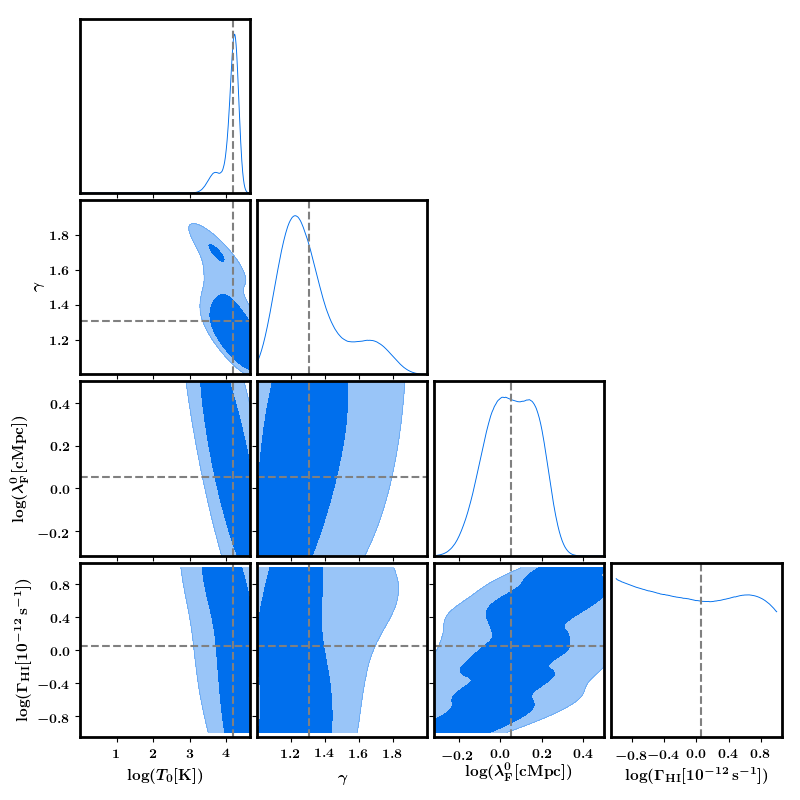}
   \caption{The likelihood contours for the analysis of the median
     line broadening distribution at $z=3.05$. The likelihood
     exploration has been done with Montepython \citep{Audren:2012wb}
     and the result plotted with GetDist utility in CosmoMC. The
     likelihood contours that we have shown in the panels refers to
     68\% and 95\% confidence level. The measured value of $T_0$ is
     anti-correlated with both $\gamma$ and $\lambda_{\rm F}$. The
     value of $\Gamma_{\rm HI}$ is correlated with $\lambda_{\rm
       F}$. The origin of these degeneracies can be understood by
     comparing to Figure~\ref{fig:trends}. The true values in the
     simulations are represented by the horizontal and vertical dashed
     lines.}
   \label{fig:contours}
\end{figure*}

In Figure~\ref{fig:contours}, we show the likelihood contours for the
parameters estimated for the redshift interval $3.0\leq z\leq
3.1$. The number of data points is 12 and the number of free
parameters is 4, hence the number of degree of freedom is 8. 
The expected anti-correlations in $\lambda_{\rm F}$-$T_0$ and
$\gamma$-$T_0$ are visible, and also the correlation between
$\Gamma_{\rm HI}$ and $\lambda_{\rm F}$.  In order to improve the
constraining power of our method, resolve the degeneracies between the
parameters, and mitigate the effect of the assumed priors on
$\Gamma_{\rm HI}$ and $\lambda_{\rm F}$, we combine the fit to the
distribution of the line broadening with the fit of the median of the
flux. We consider the analytical description of the optical depth that
we have given in Equation~\ref{eq:opticaldepth}, we can verify that it
correctly describes the median optical depth at the peak as a function
of the density contrast. We attempt to use this relation to describe
the median flux, by applying $F=\exp(-\tau)$, and considering the
distribution of density contrast as found in our reference
simulation. When we consider the median of the flux (on all the
spectra and not only on the peaks of the optical depth), we find that
Equation~\ref{eq:opticaldepth} does not match the results found in
mock spectra. This was expected, because
Equation~\ref{eq:opticaldepth} describes the relation between the
optical depth at the peak of the line and the underlying density
contrast.

\begin{figure*}
  \includegraphics[width=\textwidth]{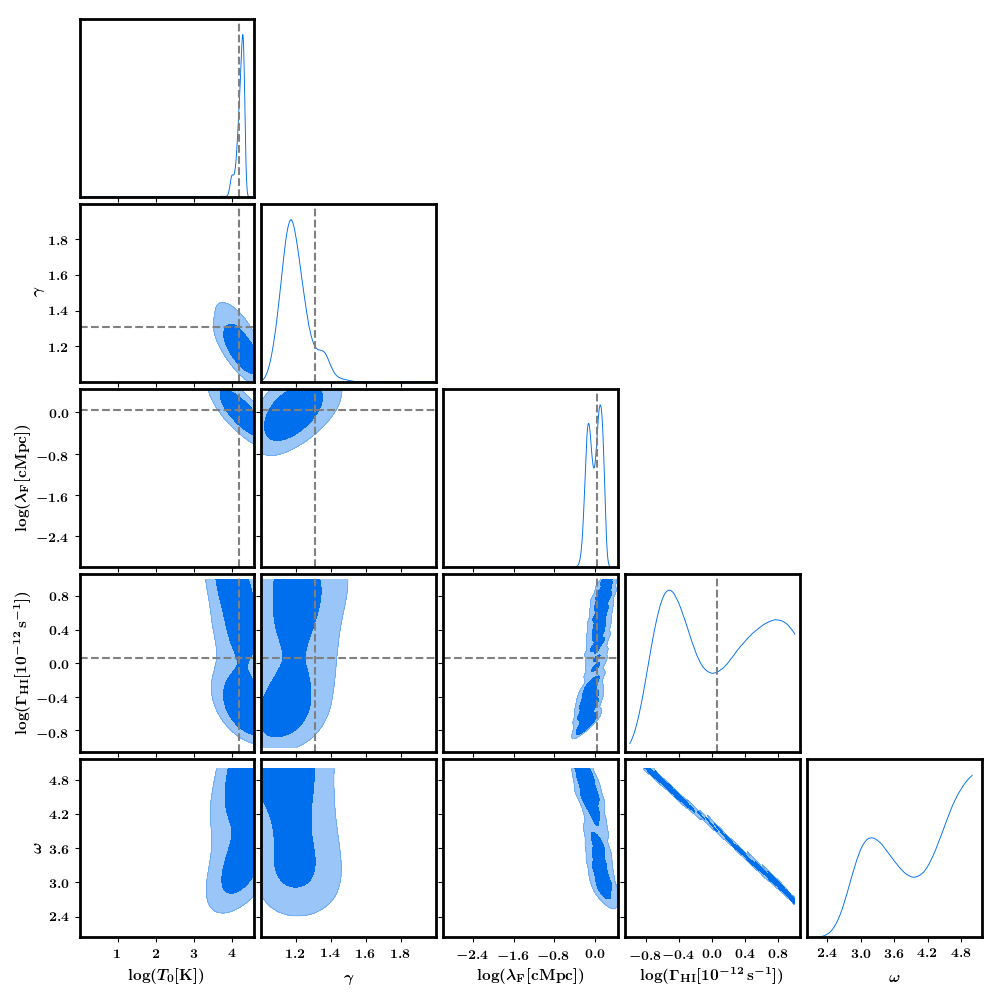}
   \caption{The likelihood contours for the joint analysis of the
     median line broadening distribution and the median
     flux at $z=3.05$, same conventions as in
     Figure~\ref{fig:contours}.  There is a slight anti-correlation
     between $T_0$ and $\lambda_{\rm F}$, a slight correlation between
     $\lambda_{\rm F}$ and $\Gamma_{\rm HI}$, and an evident
     correlation between $\Gamma_{\rm HI}$ and $\omega$. The estimated
     parameters ranges are in excellent agreement with their true
     values ($\omega$ is a nuisance parameters and it does not have an
     associated true value). }
   \label{fig:contours_joint}
\end{figure*}

\begin{figure*}
  \includegraphics[width=\columnwidth]{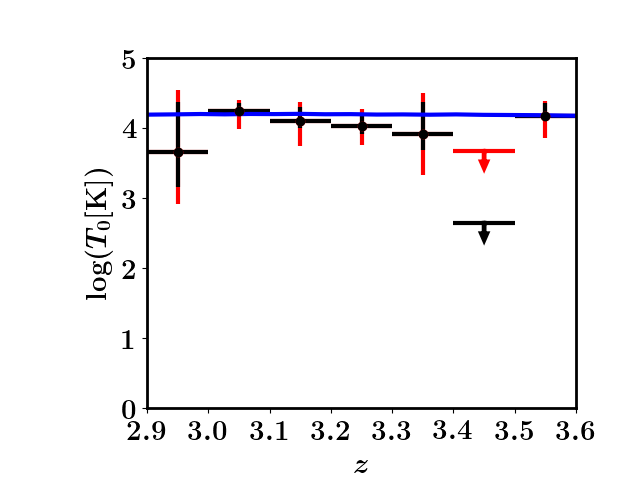}
  \includegraphics[width=\columnwidth]{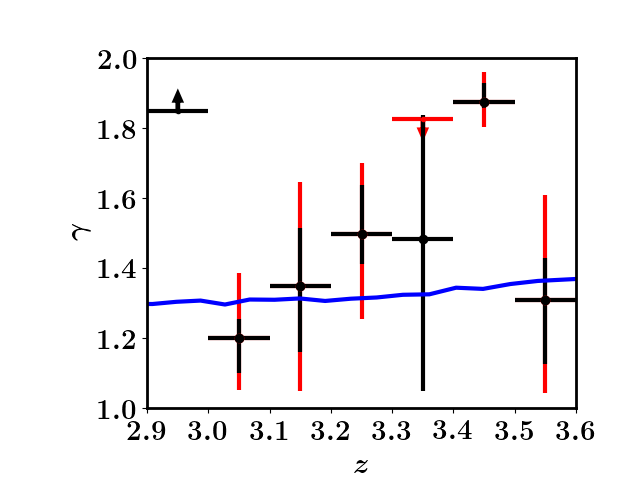}
  \includegraphics[width=\columnwidth]{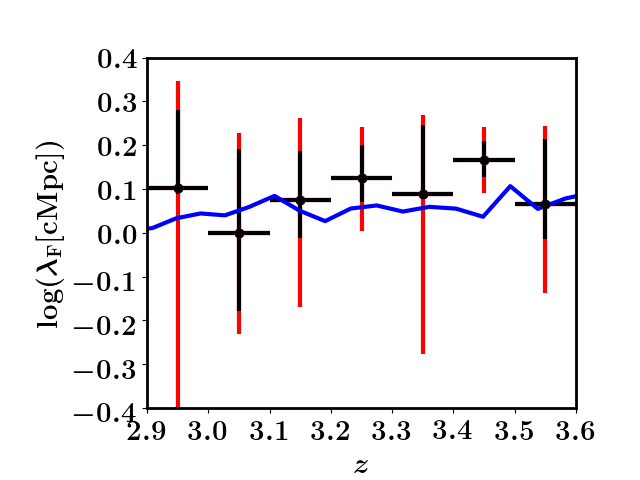}
  \includegraphics[width=\columnwidth]{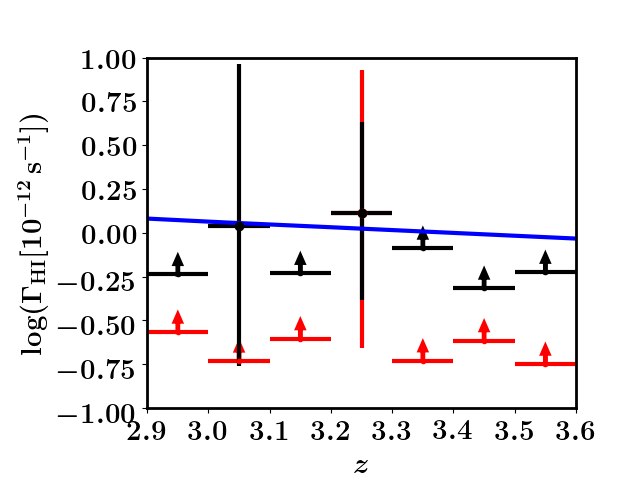}
   \caption{The result of estimation of $T_0$, $\gamma$, $\lambda_{\rm
       F}^0$ and $\Gamma_{\rm HI}$ for the analysis as in
     Figure~\ref{fig:contours_joint} for all the redshift intervals .
     The black error bars indicate the 1-$\sigma$ errors, and the red
     error bars are the 2-$\sigma$ contours.  The numbers on the top of
     each panel show the reduced chi-squared,
     $\hat{\chi^2}=\chi^2/n_{\rm dof}$, where $n_{\rm dof}$ is the number of degrees
     of freedom, that in this analysis is 13-5=8 for each redshift
     interval.
     The solid blue line are the values of the parameters
     measured from the simulations.  All the parameters are
     constrained in all the redshift bins.}
   \label{fig:nolybeta_high_taueff}
\end{figure*}

In order to account for this unknown factor, in the comparison with
the mean optical depth, we will consider the $T_0$ intervening in
\ref{eq:opticaldepth} as an additional nuisance parameter, that we
will call $\omega$. The $\chi^2$ for the joint analysis of the line
broadening and median flux will be the sum of the $\chi^2$ in
Eq.~\ref{eq:chi2} and
\begin{eqnarray}
\chi^2 &=&  \sum_i \Big(M_{\Delta}(\exp(-\tau(\Delta,\omega, \gamma,
\lambda_{\rm F}, \Gamma_{\rm HI})) \\
&- & M(F)\Big)^2/\sigma(M(F))^2 \nonumber \, ,
\end{eqnarray}
where $\tau(\Delta, \omega,\gamma, \lambda_{\rm F}, \Gamma_{\rm HI})$
is the relation between $\tau_0$ and $\Delta$ in
Eq.~\ref{eq:opticaldepth} (with the nuisance parameter $\omega$
instead of $T_0$ ), $M_{\Delta}(y(\Delta))$ represents taking the
median of the function $y$ by varying $\Delta$ over all the values of
the density contrast in the simulation, $M(F)$ is the median of $F$ as
found in the mock spectra, and $\sigma(M(F))$ is the error on the
median flux in mock spectra and it is computed by bootstrap.  In the
future application of this method to observed spectra, we will
consider the distribution of density contrast as found in our
reference simulation.  We let the nuisance parameter $\omega$ free to
vary in the interval $[0,5]$.  Now the number of data points is 13,
the number of free parameters is 5, hence the number of degree of
freedom is 8.  In Figure~\ref{fig:contours_joint}, we show the
likelihood contours for the parameters estimated for the redshift
interval $3.0\leq z\leq 3.1$.
There is an anti-correlation between $T_0$ and $\lambda_{\rm F}$,
between $\gamma$ and $T_0$, and between $\Gamma_{\rm HI}$ and
$\omega$, whereas $\gamma$ and $\lambda_{\rm F}$ are correlated.  We
show the results of this joint analysis between the line broadening
distribution and the median flux in
Figure~\ref{fig:nolybeta_high_taueff}. The parameters $T_0$,$\gamma$
and $\lambda_{\rm F}$ are detected at 2-$\sigma$ level in all the
considered redshift bins, and in excellent agreement with the true
values measured from the simulation, whereas there exist lower limits
for $\Gamma_{\rm HI}$ at 2-$\sigma$ level. 

Here, we note that the comoving size of the filaments at cosmic mean
density is $\sim 1\,{\rm cMpc}$ in all the examined redshift
intervals. This value exceeds by an order of magnitude the estimate of
the filtering length given in \citep{rorai2017}. Indeed,
\citeauthor{rorai2013} use N-body simulations for modeling the
distribution of dark matter, then they impose a smoothing filter, with
a single filtering length, for describing the baryonic density. As
discussed in \cite{schaye2000}, and as we have explicitly shown in
Figure~3 of \cite{Garzilli:2015bha}, the physical size of the
absorbers is not a single value, but it is a power-law relation of the
density. Because \citeauthor{rorai2013} does not explicitly quantify
to which density range they are more sensitive, it is not possible to
make a direct comparison with their work.

\section{Conclusions}\label{sec:conclusions}
We have described a new method to measure the IGM temperature and the
widths of the filaments that are responsible for the absorption in the
Ly$\alpha$ forest, based on the description of the minimum line
broadening that we have developed in \cite{Garzilli:2015bha} and on
the description of the median flux that we have described here. In the
original formulation, we derived a relation between the minimum line
broadening of the Ly$\alpha$ forest and the over-density,
$\Delta$. Because $\Delta$ is a quantity that cannot be measured
directly in observed quasar spectra, we reformulated the minimum line
broadening description in terms of the central line optical depth,
$\tau_0$, that can be measured directly.

In this work we considered the problem of reconstructing the line
broadening in spectra with noise and finite instrumental
resolution. We used automatic Voigt profile decompositions by VPFIT to
reconstruct noiseless spectra from noisy data, and to this
reconstructed spectra we applied the method for finding the lines and
computing the line broadening for noiseless sightlines that we
described in \cite{Garzilli:2015bha}. We have found that the 10th
percentiles of the line broadening are not very well reconstructed for
the smallest values of $\tau_0$, whereas the median line broadening is
more robust.

We applied our method to a sample of mock sightlines extracted from
our reference simulation with low and high signal to noise. Our method
is calibrated to our reference numerical simulation in two
ways. Concerning the line broadening distribution, we have determined
the multiplying factor needed to match the median line broadening to
the minimal line broadening from the reference simulation.  Concerning
the median flux, we consider the density contrast taken from
our reference simulation, and we use it to compute the observable
median flux.
We combine the analysis of the line broadening distribution with the
analysis of the median flux. We are going to discuss in an upcoming
work the application of our method to observational data.  In fact, our method
allows us to reconstruct the properties of the IGM, such as the
temperature, the size of the expanding filaments at the cosmic mean
density, and, partially, the photo-ionisation rate of neutral hydrogen.

We aim to apply this method to observed quasar spectra, in order to
obtain new measurements of the IGM temperature and of the sizes of the
absorbing structures. These measurements will be presented in a
forthcoming paper.
\section*{Acknowledgement}
The authors thanks Michele Fumagalli for reading and commenting the
earlier version of the draft. AG thanks D-ITP for supporting this
research. This work used the DiRAC Data Centric system at Durham
University, operated by the Institute for Computational Cosmology on
behalf of the STFC DiRAC HPC Facility (www.dirac.ac.uk). This
equipment was funded by BIS National E-infrastructure capital grant
ST/K00042X/1, STFC capital grants ST/H008519/1 and ST/K00087X/1, STFC
DiRAC Operations grant ST/K003267/1 and Durham University. DiRAC is
part of the National E-Infrastructure.

\newpage
\appendix
\section{A comparison with traditional Voigt profile fitting}\label{app:vpfit}
\begin{figure*}
  \includegraphics[width=\columnwidth]{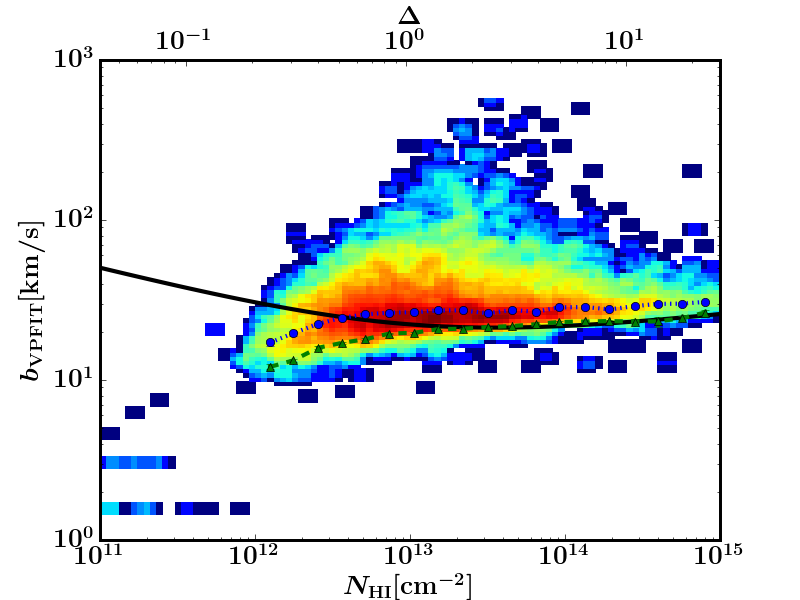}

  \caption{The distribution of line broadening as obtained from VPFIT,
    $b_{\rm VPFIT}$, versus the neutral hydrogen column density,
    $N_{\rm HI}$. The color scheme encodes the number density of the
    lines in the $b_{\rm VPFIT}$-$N_{\rm HI}$ plane. The blue dots
    (green triangles) connected by a dotted (dashed) line are the 50th
    (10th) percentiles of the $b_{\rm VPFIT}$-distribution in equally
    spaced logarithmic intervals of $N_{\rm HI}$. The solid black line
    is the line of minimum line broadening, from Eq.~(25)
    \citet{Garzilli:2015bha}. Both the 10th and the 50th percentiles
    of the $b_{\rm VPFIT}$-distribution turn towards low values of
    $b_{\rm VPFIT}$ for low values of $N_{\rm HI}$, these percentiles
    look different from the case of $b$ measured with our method,
    shown in Figure~\ref{fig:percentiles}, where the percentiles of
    the $b$ distribution turn towards high values of $b$ for low
    values of $\tau_0$, as expected theoretically.
  } 
  \label{fig:vpfitfigure}
\end{figure*}

We consider the reconstruction of percentiles of line broadening
obtained from Voigt profile fitting, which has been widely used in the
literature. Voigt profile fitting has been considered in
\citet{schaye1999,Schaye:1999vr,Ricotti:1999hx,McDonald:2000nn,bolton2012,
  Rudie:2012mx} for measuring the IGM temperature, and it is the only
line decomposition technique applied so far to the Ly$\alpha$ forest,
using a variety of codes like VPFIT, FITLYMAN \citep{fitlyman} or
AUTOVP \citep{autovp}. Voigt profile fitting is a global fitting
method that implies fitting the entire shape of the transmitted
flux. Hence, it is sensitive to the clustering of the absorbers in the
Ly$\alpha$ forest, in other words, it is sensitive to the underlying
density distribution of the gas. In fact, some Voigt profiles with
very small $b_{\rm VPFIT}$ are present, because they improve the
overall convergence of the fit. Instead, our `peak decomposition' only
measures the line broadening at ``local maxima'' in the optical
depth. We have applied Voigt profile fitting to our mock
sightlines with noise using VPFIT \citep{carswell1987,webb1987}. In
Figure~\ref{fig:vpfitfigure} we show the resulting $b_{\rm
  VPFIT}$-$N_{\rm HI}$ distribution, and compare it with the amount of
line broadening described by Eq.(~\ref{eq:bperc}).  The upturn of the
$b$-$N_{\rm HI}$ distribution that is expected for small $N_{\rm HI}$
is visible neither in the 10th nor 50th percentiles of the $b_{\rm
  VPFIT}$ distribution.

\section{Effect of the uncertainty on $\eta_{50}$}\label{app:eta50}
\begin{figure*}
  \includegraphics[width=\columnwidth]{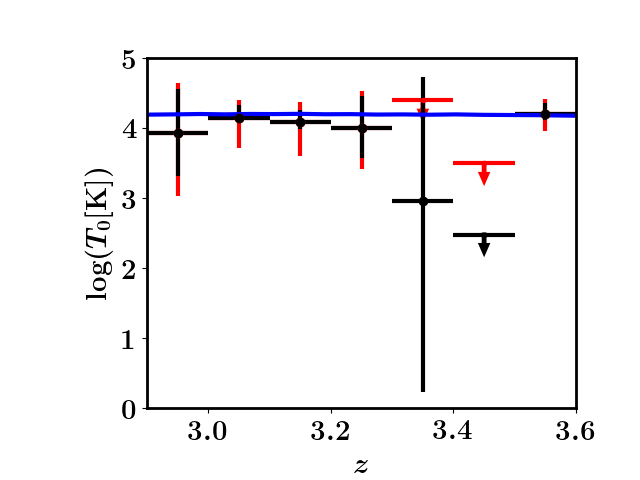}
  \includegraphics[width=\columnwidth]{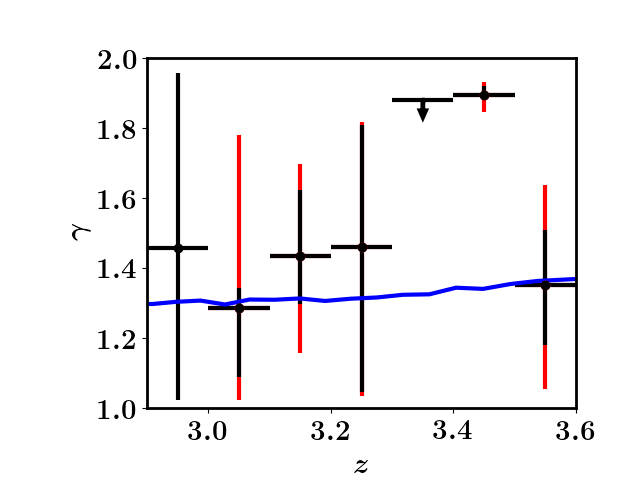}
  \includegraphics[width=\columnwidth]{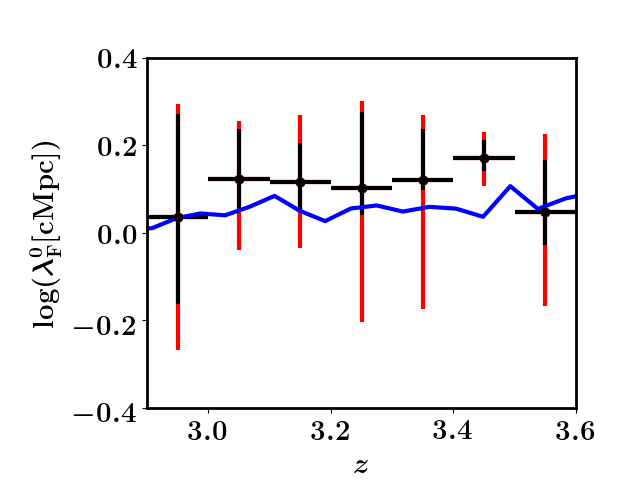}
  \includegraphics[width=\columnwidth]{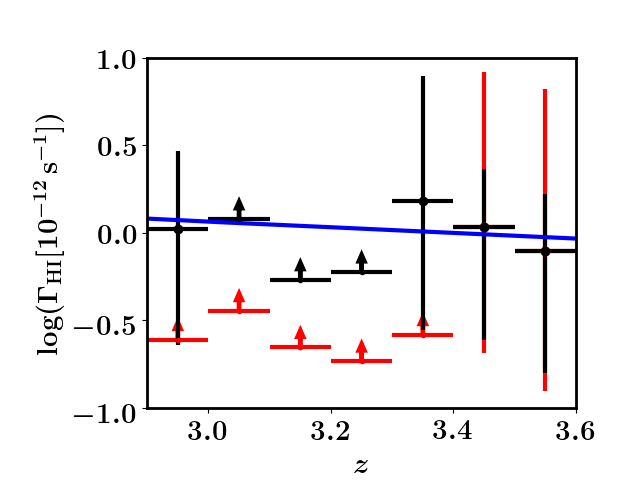}
   \caption{The results of $T_0$, $\gamma$, $\lambda_{\rm F}$ and
     $\Gamma_{\rm HI}$ estimation for a value of $\eta_{50}$ increased
     by $5\%$ respect to the value calibrated from our reference
     simulation. The analysis is performed as in
     Fig.~\ref{fig:nolybeta_high_taueff}, and we apply the same
     conventions. The result of this analysis show that our method is
     robust respect to calibration for an increase of $\eta_{50}$ of
     $5\%$.}
   \label{fig:eta50plus5}
\end{figure*}

\begin{figure*}
  \includegraphics[width=\columnwidth]{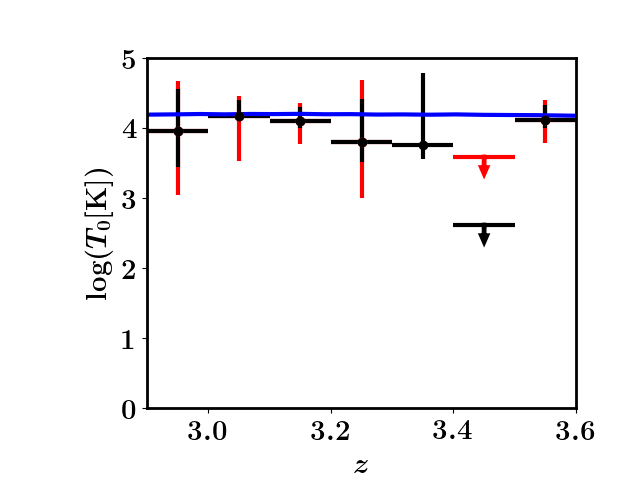}
  \includegraphics[width=\columnwidth]{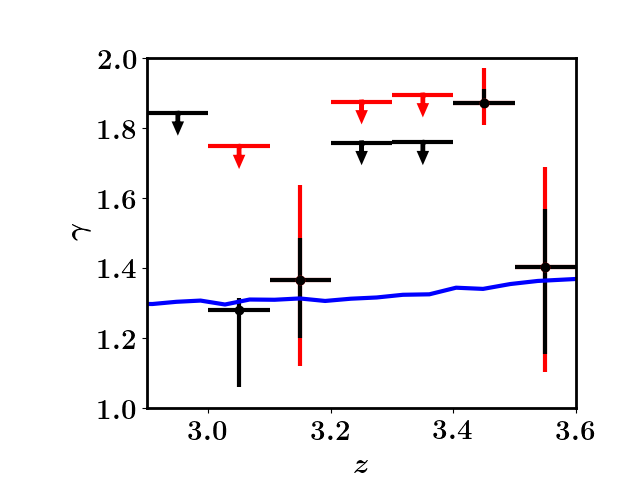}
  \includegraphics[width=\columnwidth]{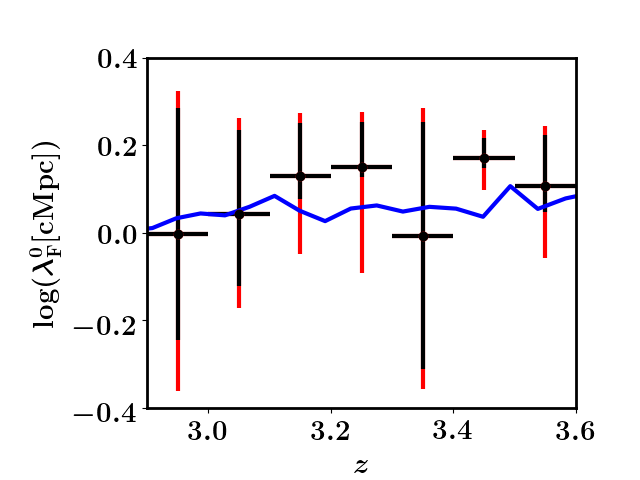}
  \includegraphics[width=\columnwidth]{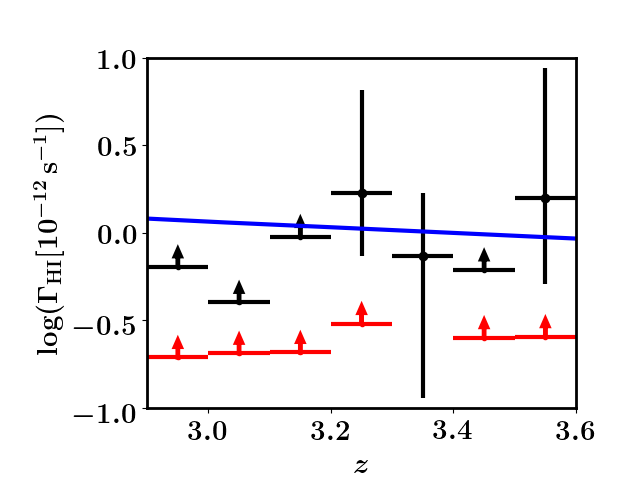}
   \caption{The results of $T_0$, $\gamma$, $\lambda_{\rm F}$ and
     $\Gamma_{\rm HI}$ estimation for a value of $\eta_{50}$ decreased
     by $5\%$ respect to the value calibrated from our reference
     simulation. The analysis is performed as in
     Fig.~\ref{fig:nolybeta_high_taueff}, and we apply the same
     conventions. The result of this analysis show that our method is
     robust respect to calibration for an decrease of $\eta_{50}$ of
     $5\%$. }
   \label{fig:eta50minus5}
\end{figure*}

We show that changing the values of $\eta_{50}$ by $\div 5\%$ does not
affect the result of our analysis. In Figure~\ref{fig:eta50plus5}
(Figure~\ref{fig:eta50minus5}) we show the constraints on the IGM
parameters for the case that $\eta_{50}$ is increased (decreased) by
$5\%$. We infer that a variation of $\eta_{50}$ within $5\%$ does not
affect the result of our analysis.

\section{Considering  lower S/N}\label{app:lowsn} 
\begin{figure*}
  \includegraphics[width=\columnwidth]{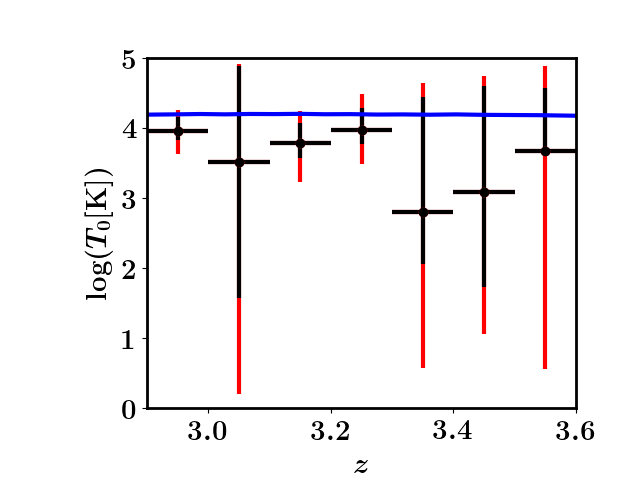}
  \includegraphics[width=\columnwidth]{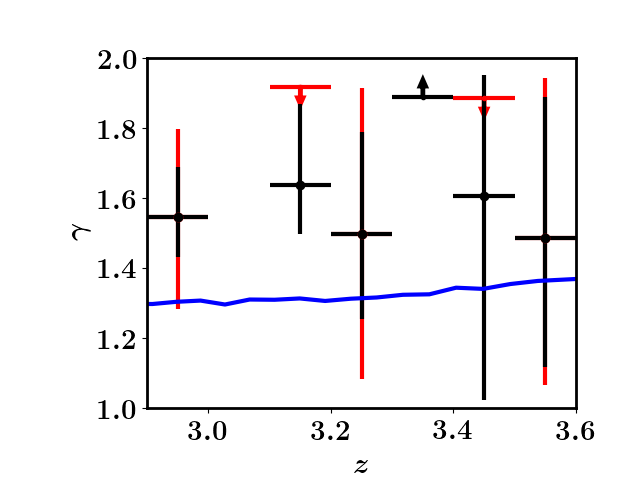}
  \includegraphics[width=\columnwidth]{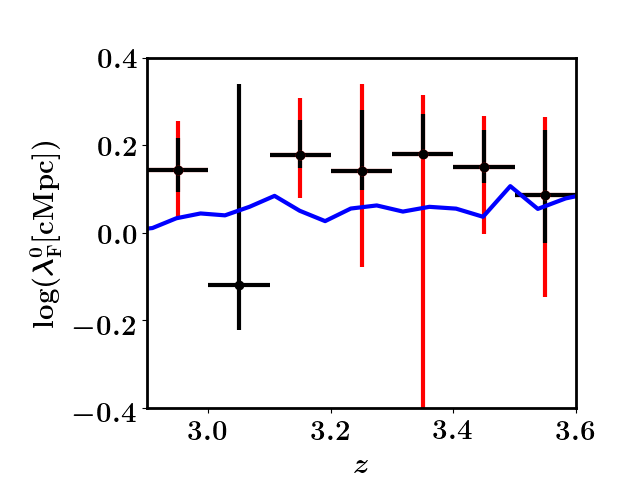}
  \includegraphics[width=\columnwidth]{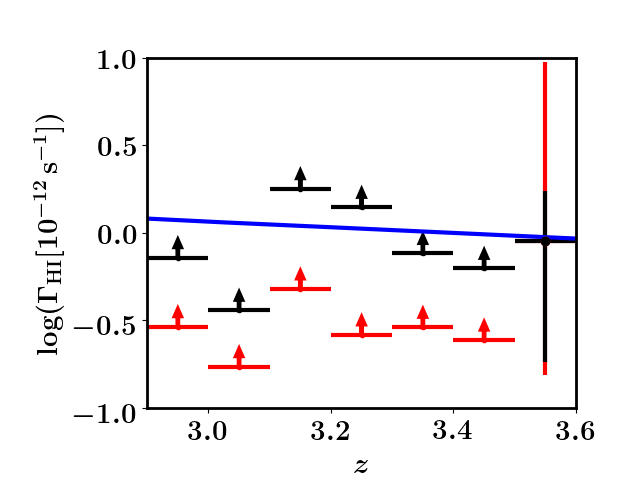}
   \caption{The results of $T_0$, $\gamma$, $\lambda_{\rm F}$ and
     $\Gamma_{\rm HI}$ estimation for the case of lower
     signal-to-noise spectra (S/N=30), each redshift bin is analyzed
     independently. The $b$-$\tau_0$ relation is fitted in the
     interval $\tau_0 \in [0.3,4]$. There are no constraints on
     $\Gamma_{\rm HI}$. Same conventions as in
     Fig.~\ref{fig:nolybeta_high_taueff}. }
   \label{fig:nolybeta_low}
\end{figure*}

In Figure~\ref{fig:nolybeta_low} we show the results of the parameters
estimation for the case of a low signal-to-noise sample of spectra
(S/N=30) for the central optical depth interval $\tau_0\in [0.3, 4]$
(that is different from the optical depth interval that we have chosen
for the high signal-to-noise sample). The results are similar to the
ones found in the high signal-to-noise case, but with larger error
bars.

We conclude that our method also works with lower signal to noise
spectra, and it is hence applicable to existing spectra.

\bibliographystyle{mnras} \bibliography{main,main2}

\appendix
\end{document}